\documentclass[apj]{emulateapj}
\usepackage{subfigure}
\usepackage{graphics}

\def\spose#1{\hbox to 0pt{#1\hss}}
\def\simlt{\mathrel{\spose{\lower 3pt\hbox{$\mathchar"218$}}
    \raise 2.0pt\hbox{$\mathchar"13C$}}}
\def\simgt{\mathrel{\spose{\lower 3pt\hbox{$\mathchar"218$}}
    \raise 2.0pt\hbox{$\mathchar"13E$}}}

\newcommand{\oiiin}{\mbox{[\ion{O}{3}]}}
\newcommand{\oiii}{\mbox{[\ion{O}{3}]} $\,$}
\newcommand{\oiiiw}{\mbox{[\ion{O}{3}] $\lambda$5007} $\,$}
\newcommand{\oiiiwn}{\mbox{[\ion{O}{3}] $\lambda$5007}}
\newcommand{\oiiiww}{\mbox{[\ion{O}{3}] $\lambda$4959} $\,$}

\newcommand{\hb}{\mbox{H$\beta$} $\,$}
\newcommand{\hbn}{\mbox{H$\beta$}}
\newcommand{\ha}{\mbox{H$\alpha$} $\,$}
\newcommand{\han}{\mbox{H$\alpha$}}

\newcommand{\nii}{\mbox{[\ion{N}{2}] $\lambda$6584} $\,$}
\newcommand{\niin}{\mbox{[\ion{N}{2}] $\lambda$6584}}
\newcommand{\niiw}{\mbox{[\ion{N}{2}] $\lambda$6548} $\,$}
\newcommand{\niiwn}{\mbox{[\ion{N}{2}] $\lambda$6548}}
\newcommand{\ntwo}{\mbox{[\ion{N}{2}]} $\,$}

\newcommand{\siiha}{\mbox{[\ion{S}{2}] $\lambda\lambda$6716,31}/{\mbox{H$\alpha$} $\,$}}
 
\newcommand{\oiha}{\mbox{[\ion{O}{1}] $\lambda$6300}/{\mbox{H$\alpha$} $\,$}}

\newcommand{\oi}{\mbox{[\ion{O}{1}]} $\,$}

\newcommand{\sii}{\mbox{[\ion{S}{2}]} $\,$}

\slugcomment{\it Accepted for Publication in ApJ}
\shortauthors{Comerford, Schluns, Greene, \& Cool}
\shorttitle{Dual Supermassive Black Hole Candidates in the AGN and Galaxy Evolution Survey}

\begin{document}

\title{Dual Supermassive Black Hole Candidates in the AGN and Galaxy Evolution Survey}

\author{Julia M. Comerford}
\affil{Department of Astrophysical and Planetary Sciences, University of Colorado, Boulder, CO 80309, USA}
\affil{Astronomy Department, University of Texas at Austin, Austin, TX 78712, USA}

\author{Kyle Schluns}
\affil{Department of Astronomy, Boston University, 725 Commonwealth Avenue, Boston, MA 02215, USA}

\author{Jenny E. Greene}
\affil{Department of Astrophysical Sciences, Princeton University, Princeton, NJ 08544, USA}

\and

\author{Richard J. Cool}
\affil{MMT Observatory, Tucson, AZ 85721-0065, USA} 

\begin{abstract}
Dual supermassive black holes (SMBHs) with kiloparsec scale separations in merger-remnant galaxies are informative tracers of galaxy evolution, but the avenue for identifying them in large numbers for such studies is not yet clear.  One promising approach is to target spectroscopic signatures of systems where both SMBHs are fueled as dual active galactic nuclei (AGNs), or where one SMBH is fueled as an offset AGN.  Dual AGNs may produce double-peaked narrow AGN emission lines, while offset AGNs may produce single-peaked narrow AGN emission lines with line-of-sight velocity offsets relative to the host galaxy.  We search for such dual and offset systems among 173 Type 2 AGNs at $z<0.37$ in the AGN and Galaxy Evolution Survey (AGES), and we find two double-peaked AGNs and five offset AGN candidates.  When we compare these results to a similar search of the DEEP2 Galaxy Redshift Survey and match the two samples in color, absolute magnitude, and minimum velocity offset, we find that the fraction of AGNs that are dual SMBH candidates increases from $z=0.25$ to $z=0.7$ by a factor of $\sim6$ (from 2/70 to 16/91, or $2.9^{+3.6}_{-1.9}\%$ to $18^{+5}_{-5}\%$).  This may be associated with the rise in the galaxy merger fraction over the same cosmic time.  As further evidence for a link with galaxy mergers, the AGES offset and dual AGN candidates are tentatively $\sim3$ times more likely than the overall AGN population to reside in a host galaxy that has a companion galaxy (from 16/173 to 2/7, or $9^{+3}_{-2}\%$ to $29^{+26}_{-19}\%$).  Follow-up observations of the seven offset and dual AGN candidates in AGES will definitively distinguish velocity offsets produced by dual SMBHs from those produced by narrow-line region kinematics, and will help sharpen our observational approach to detecting dual SMBHs.
\end{abstract}


\section{Introduction}
\label{intro}

A merger between two galaxies, each with its own central supermassive black hole (SMBH), results in a merger-remnant galaxy hosting two SMBHs.   As these SMBHs orbit in the potential of the host galaxy, they are dragged closer together by dynamical friction from the surrounding stars, form a gravitationally-bound binary system, and ultimately merge \citep{BE80.1,MI01.1}.  While the SMBHs are at kiloparsec (kpc) scale separations, before they are bound as a binary system, they are known as dual SMBHs.  Gas churned up by the galaxy merger may accrete onto the dual SMBHs, and cases where one or both of the SMBHs fuel active galactic nuclei (AGNs) are known as offset AGNs and dual AGNs, respectively. These offset and dual AGNs, which we refer to here as having separations $<10$ kpc, have valuable potential as new observational tools for studies of galaxy evolution, including measurements of galaxy merger rates, SMBH mass growth, and SMBH merger rates.

However, because of their small spatial separations from the host galaxy centers, offset and dual AGNs are challenging to identify.  As a consequence, early discoveries of offset AGNs and dual AGNs were serendipitous \citep{KO03.1,BA08.1}. While ultra-hard X-rays have been used to uncover a handful of dual AGNs \citep{KO11.1,KO12.1}, serendipitous discoveries of dual AGN candidates continue today (e.g., \citealt{CO09.3,FA11.1,BA12.1}). To accelerate the discovery rate of offset and dual AGNs, a new systematic approach was developed to identify candidates in spectroscopic surveys of galaxies.  This technique uses galaxy spectra to select offset and dual AGN candidates as narrow AGN emission lines with line-of-sight velocity offsets of a few hundred km s$^{-1}$ relative to the host galaxy stellar absorption features.  Offset AGN candidates display velocity offsets in single-peaked narrow AGN emission lines relative to stellar absorption lines, while the dual AGN candidates have double-peaked narrow lines. Numerical simulations of AGN activity in galaxy mergers show that these double-peaked lines are indeed useful selectors of dual AGNs \citep{VA12.1,BL13.1}.  

The velocity-offset approach was first applied systematically to the DEEP2 Galaxy Redshift Survey, where it was used to identify 30 offset AGN candidates \citep{CO09.1} and two dual AGN candidates \citep{GE07.1,CO09.1} at $0.35<z<0.79$. Subsequently, it was used in the Sloan Digital Sky Survey (SDSS) to uncover 340 unique double-peaked AGNs at $0.01<z<0.69$ \citep{WA09.1,LI10.1,SM10.1} and 131 double-peaked quasars at $0.8 < z < 1.6$ \citep{BA13.1}.  These double-peaked AGNs have been a reservoir for follow-up studies aimed at distinguishing line profiles produced by dual AGNs from those produced by narrow-line region kinematics (e.g., \citealt{LI10.2,CO11.2,GR11.1,MC11.1,RO11.1,TI11.1,CO12.1,FU12.1}).  Some of these observations have resulted in confirmations of dual AGNs \citep{FU11.3,LI13.1}, but the majority of double-peaked AGNs are likely produced by gas kinematics associated with a single AGN (e.g., \citealt{FU11.1,SH11.1}).

Given the successes of using velocity-offset AGN emission lines to select offset and dual AGN candidates in DEEP2 and SDSS, it is a natural extension to apply the same approach to other large spectroscopic surveys of galaxies, such as the AGN and Galaxy Evolution Survey (AGES).  
A search for offset and dual AGN candidates in AGES, at a mean redshift of $\bar{z}=0.25$, would fill the gap between the lower redshift SDSS ($\bar{z}=0.1$) and higher redshift DEEP2 ($\bar{z}=0.7$) samples of dual SMBH candidates, enabling dual SMBHs to be used for studies of galaxy evolution across the full range of $0 \simlt z \simlt 1$.

Here we present the results of our systematic search for velocity-offset narrow AGN emission lines in AGES galaxy spectra, which uncovered five offset AGN candidates and two dual AGN candidates.  These galaxies, at $0.08 < z < 0.36$, are promising candidates for follow-up observations that will definitively determine whether they host offset and dual AGNs.

The remainder of this paper is organized as follows.  In Section 2, we describe the AGES spectra, AGN selection, and our measurements of the redshifts, emission line fluxes, and velocity offsets of the emission lines.  In Section 3, we present our identifications of two double-peaked AGNs and five offset AGN candidates, examine their host galaxies, and compare them to similar candidates in other large spectroscopic surveys of galaxies.  Section 4 gives our conclusions.  We assume a Hubble constant $H_0 =70$ km s$^{-1}$ Mpc$^{-1}$, $\Omega_m=0.3$, and $\Omega_\Lambda=0.7$ throughout, and all distances are given in physical (not comoving) units.

\section{The Sample and Analysis}

\subsection{The Sample}

Our sample consists of a catalog of optical galaxy spectra observed for AGES \citep{KO12.3,CO12.2}.  Using Hectospec, an optical fiber-fed spectrograph with $1\farcs5$ fibers on the MMT 6.5 m telescope, AGES observed 7.7 deg$^2$ of the Bo\"{o}tes field in the NOAO Deep Wide-Field Survey \citep{JA99.2}. The resultant spectra have a wavelength coverage of 3700 -- 9200 \AA, and the spectral resolution is 6 \AA, yielding $R \sim 1000$.  AGES determined spectroscopic redshifts for 18,163 galaxies to a limiting magnitude of $I=20$.

Since we will use the \hbn, \oiiiwn, \han, and \nii emission lines to diagnose AGN activity (see Section~\ref{agn}), we select the galaxy spectra where all four of these emission lines are within the AGES wavelength range.  This cut results in 8136 spectra at $z < 0.37$, and this sample is the focus of our analysis as described below.

\subsection{Host Galaxy Redshift Measurements} 
\label{stellar}

While redshifts for the AGES galaxies have already been measured by cross correlation with emission and absorption line galaxy and AGN template spectra \citep{KO12.3}, these redshifts may be weighted towards the emission lines and hence not true representations of the stellar absorption redshifts.  Our selection of velocity-offset emission lines depends on the redshift of the galaxy's stellar absorption features, so we measure these redshifts using the high equivalent width absorption lines Ca H+K, G-band, and \ion{Mg}{1} {\it b}.

For each of the 8136 spectra in our sample, we constructed a rest-frame, model spectrum of the stellar continuum with the GANDALF algorithm \citep{SA06.1}, and from the template we isolated a region of flux, covering 100 \AA, around each of the three stellar absorption features. We fit a redshift to each of the three regions, and we took the host galaxy redshift to be the mean of these three redshifts.

\subsection{Emission Line Flux Measurements}
\label{emission}

We measured the fluxes of nine emission lines (\hbn; \oiii $\lambda\lambda4959, 5007$; \oi $\lambda6300$; \han; \ntwo $\lambda\lambda6548, 6584$; \sii $\lambda\lambda6716, 6731$) covered in the wavelength range of our sample. After subtracting a continuum model from each spectrum, we fit Gaussians to each emission feature. We fixed the values of \oiiiwn/\oiiiww and \niin/\niiw according to the line flux ratios set by atomic physics, and we tied the wavelength centroids of \hb and \oiii together and the wavelength centroids of \ha and \ntwo together.  We also tied the widths of the \hbn, \oiiin, \han, and \ntwo profiles. 

We fit both single and double Gaussian profiles to each emission line, since double Gaussian fits are more suitable for double-peaked profiles or systems with wide emission-line wings.  For the double Gaussian fits, we applied all of the criteria described above for the primary (narrower) Gaussian, while for the secondary (broader) Gaussian we tied the wavelength centroids to each other, the Gaussian widths to each other, and the ratios of the primary Gaussian flux to the secondary Gaussian flux to each other. 

We defined a double Gaussian fit as more appropriate for a spectrum when the following relation \citep{HA05.1} applies for all six emission lines:
	
\begin{equation}
\frac{\chi^2 _{\mathrm{single}} - \chi^2 _{\mathrm{double}}}{\chi^2 _{\mathrm{double}}} > 0.2 \; ,
\label{chi-squared}
\end{equation}
where $\chi^2_{\mathrm{single}}$ and $\chi^2_{\mathrm{double}}$ are the chi-squared values for the single Gaussian fits and the double Gaussian fits, respectively, to the emission lines.

We measured the emission line fluxes as the areas under the best-fit Gaussians, with uncertainties propagated from the errors on the fitting parameters.

\begin{deluxetable*}{ccccccccc}
\tabletypesize{\scriptsize}
\setlength{\tabcolsep}{0.01in} 
\tablecaption{Observed Properties of AGES AGNs}
\tablewidth{0pt}
\tablecolumns{9}
\tablehead{ 
  \colhead{ID}  & 
  \colhead{Host }   & 
  \colhead{\hb} & 
  \colhead{\oiiiw} & 
  \colhead{\ha} & 
  \colhead{\oiiiwn/} & 
  \colhead{\niin/} &
  \colhead{\mbox{[\ion{S}{2}] $\lambda\lambda$6716,31}/} &
   \colhead{\mbox{[\ion{O}{1}] $\lambda$6300}/} \\
  \colhead{} &
  \colhead{Galaxy} &
  \colhead{Velocity Offset} &
  \colhead{Velocity Offset} &
  \colhead{Velocity Offset} &
  \colhead{\hb} &
  \colhead{\ha} &
  \colhead{\ha} &
  \colhead{\ha}  \\
  \colhead{} &
  \colhead{Redshift$^a$} &
  \colhead{(km s$^{-1}$)} &
  \colhead{(km s$^{-1}$)} &
  \colhead{(km s$^{-1}$)} &
  \colhead{} &
    \colhead{} &
      \colhead{} &
   \colhead{}}
\startdata
{\tiny J142445.51+331438.8} & 0.2675 & $-90.7 \pm 27.3$ & \phn \phd $50.9 \pm 18.7$ &  $-84.5 \pm 18.3$ & $4.36 \pm 0.58$ & $0.84 \pm 0.11$ & $0.41 \pm 0.21$ & $0.08 \pm 0.03$ \\
{\tiny J142452.14+343209.4} & 0.2613 & $-14.3 \pm 32.5$ &  \phn \phd $51.7 \pm 12.7$ & $-20.2 \pm 26.5$ & $6.77 \pm 0.91$ & $0.73 \pm 0.11$ & $0.80 \pm 0.33$ & $0.63 \pm 0.22$ \\
{\tiny J142505.76+324733.2} & 0.3694 & $-49.9 \pm 33.5$ & $-40.8 \pm 22.2$ & $-82.0 \pm 33.9$ & $2.00 \pm 0.22$ & $0.78 \pm 0.21$ & $0.01 \pm 0.03$ & $0.01 \pm 0.01$ \\
{\tiny J142506.00+350927.5} & 0.1921 & $-24.1 \pm 18.4$ &  \phn $-2.6 \pm 12.5$ & $-38.6 \pm 15.1$ & $3.17 \pm 0.22$ & $0.49 \pm 0.04$ & $0.31 \pm 0.18$ & $0.15 \pm 0.14$ 
\enddata
\label{tbl:agn}
\tablecomments{$^a$The uncertainties on the host galaxy redshifts are 0.0003. \\ (This table is available in its entirety in a machine-readable form in the online journal. A portion is shown here for guidance regarding its form and content.)}
\end{deluxetable*}

\subsection{Quality Criteria}
\label{quality}

Since we will use the \hbn, \oiiiwn, \han, and \nii emission lines to select AGNs (see Section~\ref{agn}), it is essential to a clean selection that these emission lines are significantly detected. To remove noisy spectra, we required that each of the four emission lines is detected with at least 2$\sigma$ significance. We also required an \ha equivalent width $>$ 5 \AA, where we measured equivalent width as the ratio of the \ha flux to the median flux in the continuum near \han.

AGES observed some galaxies more than once, creating duplicate spectra in the catalog.  In cases where both duplicate spectra passed the quality cuts described above, we retained the spectrum with the smaller chi-squared value for the fit to the emission lines (Section~\ref{emission}) for the rest of our analysis.  After the quality cuts and removal of duplicates, our sample consists of high quality spectra of 4481 galaxies, or 55$\%$ of the original catalog at $z<0.37$.

\subsection{AGN Selection Criteria}
\label{agn}

\begin{figure}
\begin{center}
\includegraphics[width=9cm]{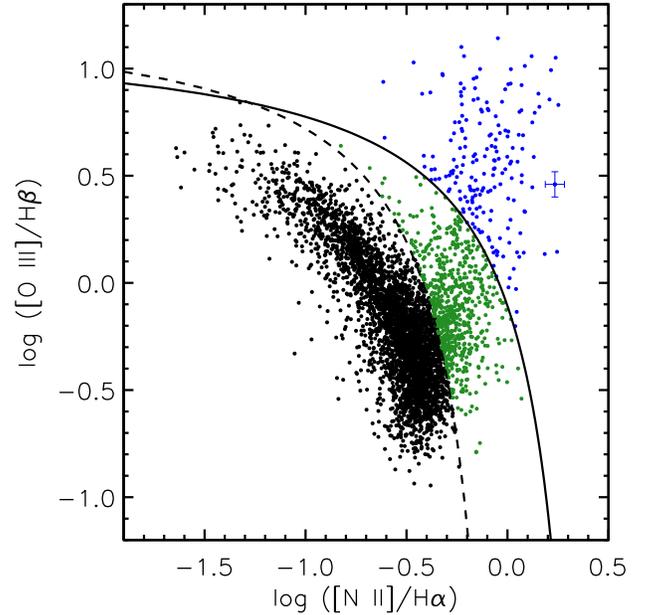}
\end{center}
\caption{BPT diagnostic diagram used to identify AGNs in the quality-selected sample of 4481 galaxy spectra in AGES. The dashed line illustrates the empirical division between galaxies that are purely star-forming and those that are dominated by a combination of star formation and AGN activity \citep{KA03.1}, while the solid line shows the theoretical maximum for starbursts \citep{KE01.2}.  The blue points show the pure AGNs, the green points show the AGN -- starburst composites, and the black points show the purely star-forming galaxies. For illustrative purposes, the median error bars are shown on one data point.}
\label{fig:BPT}
\end{figure}

Using our measured emission line fluxes (Section~\ref{emission}), we selected AGNs with the standard Baldwin--Phillips--Terlevich (BPT) diagram of line ratios (\citealt{BA81.1,VE87.1,KE06.1}; Figure~\ref{fig:BPT}). Our selection of galaxies with line flux ratios above the theoretical AGN -- starburst boundary \citep{KE01.2} yielded 182 AGN candidates.  We inspected these 182 candidates for quality by eye and found that one has poorly subtracted night sky lines and eight have artificially rising fluxes towards the red end of the spectra, which call into question the accuracy of the fits to their line fluxes and subsequent selection as AGNs. After removing these nine objects, our sample consists of 173 AGNs (Table~\ref{tbl:agn}).  All of the AGNs are Type 2 AGNs, since the \hb and \ha emission lines are well reproduced by a model that ties their widths to the \oiii and \ntwo line widths without an additional broad component (Section~\ref{emission}).

 The \siiha and \oiha line flux ratios can also be used to distinguish Seyferts, low-ionization narrow emission-line regions (LINERs) where the power sources are still unclear (e.g., \citealt{HE80.1,SH92.1,DO95.1,AL00.1,HO08.2}), and ambiguous cases that do not have uniform classifications across the line diagnostics.   The line flux ratios of the 173 AGNs show that 82 are Seyferts, 22 are LINERs, and 69 are ambiguous \citep{VE87.1,KE06.1}.  Of the 69 ambiguous cases, 26 have line flux ratios that are, within their errors, consistent with Seyfert classifications.  Therefore, a total of 108 AGNs ($62^{+4}_{-3} \%$ of the population) are either classified as Seyferts or ambiguous systems that are consistent with Seyfert classifications.
 
We proceed with our analysis of the 173 AGNs, noting that the majority of the AGES AGN are likely Seyferts and that the other studies to which we will compare (Section~\ref{compare}) used AGN classifications without separations for LINERs and ambiguous cases.

\subsection{More Rigorous Fits to the AGN Emission Lines}
\label{rigorous}

For this smaller sample of 173 Type 2 AGNs we redid the fits to the continuum-subtracted emission lines and allowed the velocities of the emission lines to float relative to each other, since in principle different lines may display different redshifts (e.g., \citealt{FI84.1,BO02.1,MU08.1,LU12.1}).
As in Section~\ref{emission}, we applied both single and double Gaussian fits to the lines. 
In the case of a single Gaussian or a primary (narrower) component of a double Gaussian, we set the wavelength centroids of the \hbn, \oiiin, and \ha features to be free parameters, while we tied the \ntwo wavelength centroids to that of \han. We also tied the wavelength centroids of all of the secondary (broader) Gaussians to one another.  The median full widths at half maximum of the best fit narrow and broad components are 6 \AA \, and 14 \AA, respectively.

We used the chi-squared criterion (Equation~\ref{chi-squared}) to determine whether a single or double Gaussian is the more suitable fit for each spectrum.  In some instances we found that a double Gaussian fit is appropriate for the \oiii lines while a single Gaussian fit is appropriate for the \hbn, \han, and \ntwo lines, as has been seen in other AGN samples (e.g., \citealt{HO97.1,GR05.1}).

\begin{figure}
\begin{center}
\includegraphics[width=9cm]{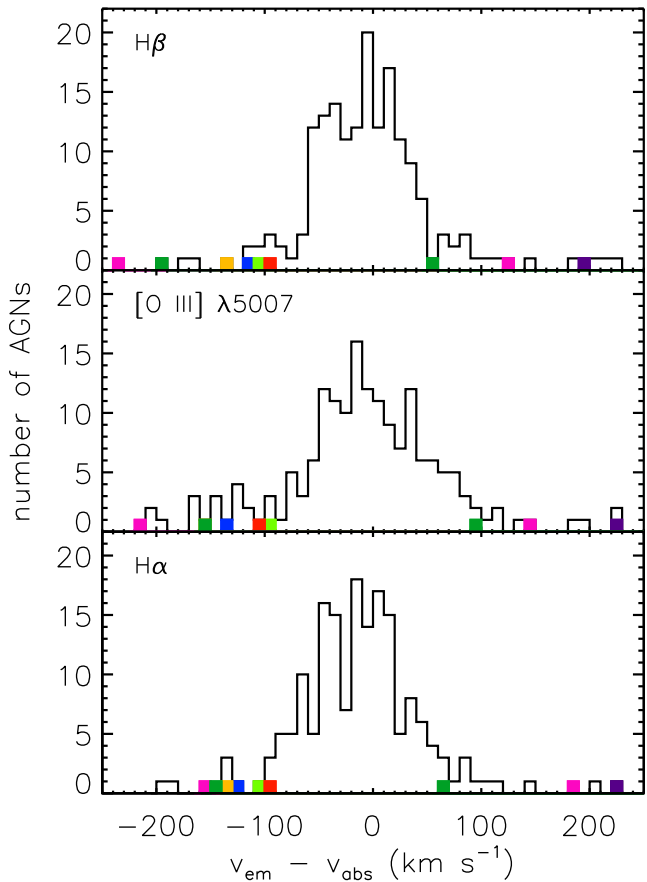}
\end{center}
\caption{Histograms of the line-of-sight velocity offsets of \hb (top), \oiiiw (middle), and \ha (bottom) relative to the host galaxy stellar absorption features for our sample of 173 Type 2 AGNs in AGES. The colored segments denote the velocity offsets of the five offset AGN candidates and the two dual AGN candidates, as selected in Section~3 (see Figure~\ref{fig:offset} for color identifications). The velocity offsets of both the redward and blueward peaks of the double-peaked AGNs are shown.}
\label{fig:hist}
\end{figure}

\subsection{Flux and Velocity Offset Measurements for AGN Emission Lines}
\label{agnmeasure}

To determine the robustness of the fits to the emission lines, we added noise to each AGN spectrum and redid the fits described in Section~\ref{rigorous}.  Using Monte Carlo realizations, we drew the noise 100 times from a Gaussian distribution with the variance of the spectrum's flux and redid the emission line fits each time. The noise is dominated by variance in the sky subtraction, while systematics in flat-fielding and flux calibration are unknown and not accounted for here.
For each realization we measured the flux of each emission line as the area under the Gaussian fit, and we used the mean and standard deviation of these values as the line's flux and uncertainty.  We also used the host galaxy redshifts to convert these line fluxes to luminosities.

The Gaussian fit to each line also yields a wavelength centroid, and in the case of a double Gaussian fit we used the central wavelength of the primary (narrower) Gaussian as the line's representative wavelength.  We measured the wavelength of each emission line as the mean and standard deviation of the wavelength centroids determined from the 100 realizations of adding noise to each spectrum.      

We then combined the wavelengths of the emission lines and the wavelengths of the stellar absorption features (measured in Section~\ref{stellar}) to determine the line-of-sight velocity offset, $v_{em}-v_{abs}$, of each AGN emission line relative to the host galaxy stellar absorption features.  When we compare the velocity offsets of the emission lines in each galaxy, we find that the median velocity difference of \hb (\han) relative to \oiiiw is $2 \pm 40$ km s$^{-1}$ ($-7 \pm 30$ km s$^{-1}$). Figure~\ref{fig:hist} shows histograms of the velocity offsets $v_{em}-v_{abs}$ measured for \hbn, \oiiiwn, and \han.  To obtain an overall averaged velocity offset for each system, we took the mean of the three velocity offset measurements weighted by their inverse variances. 

\section{Results}

\subsection{Identification of Two Double-peaked AGNs}
\label{dp}

\begin{figure}
\begin{center}
\includegraphics[width=9cm]{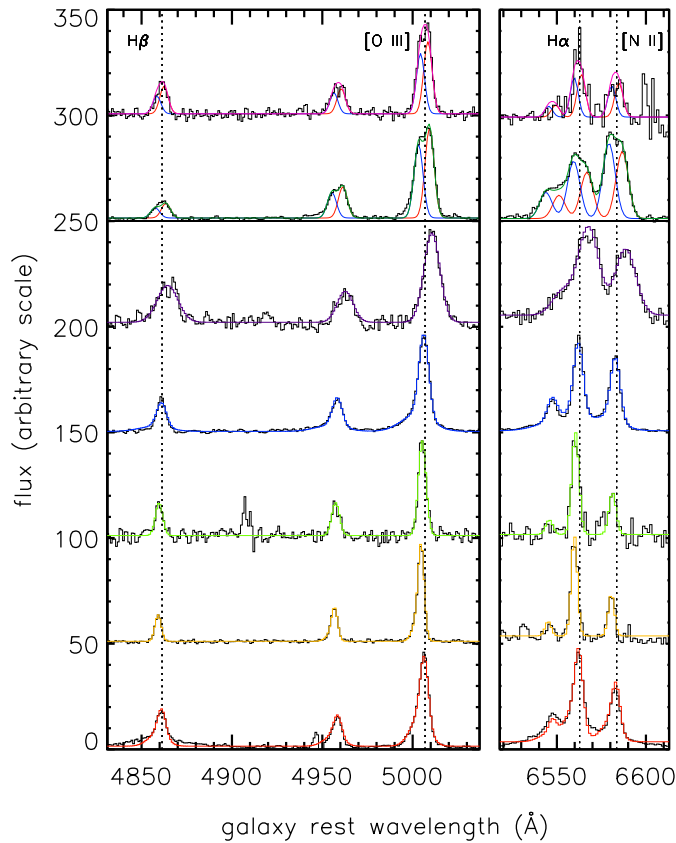}
\end{center}
\caption{Segments of the AGES spectra of the two double-peaked AGNs (top) and five offset AGN candidates (bottom), where the best fits to the spectra are shown as the colored curves.  The candidates are presented in the same order, from top to bottom, as they are listed in Table~\ref{tbl:properties}. For clarity, the spectra are normalized in flux and offset vertically.  Each spectrum is shifted to the rest wavelength of its host galaxy, based on the host galaxy redshift given by the stellar absorption features.  The dotted vertical lines show the restframe wavelengths of \hbn, \oiiiwn, \han, and \niin.  The blueward and redward components of the double Gaussian fits to the double-peaked AGNs are shown as the blue and red curves, respectively, and the emission line velocity shifts are also visible in the offset AGN candidates.
}
\label{fig:offset}
\end{figure}

We examined the 173 AGN spectra by eye and selected 20 spectra as candidates for double-peaked emission lines.  To determine which spectra are truly double-peaked, and hence dual AGN candidates, we fit double Gaussian profiles to the emission lines in each of the 20 spectra.  We selected the double-peaked AGNs as those that have \oiiiwn/\hb and \niiwn/\ha flux ratios that indicate that both the redward and the blueward emission components are produced by AGNs \citep{BA81.1,VE87.1,KE06.1}, which yielded two double-peaked AGNs (Figure~\ref{fig:offset}; Table~\ref{tbl:properties}). 

Using the inverse-variance-weighted mean velocity offsets of the red peaks and blue peaks, we find the line-of-sight velocity separations of the two double-peaked AGNs are $358.1 \pm 4.3$ km s$^{-1}$ for NDWFS J143208.27+353255.5 and $246.1 \pm 37.4$ km s$^{-1}$ for NDWFS J143359.71+351020.5.  At the redshifts of these galaxies, the minimum velocity separations resolvable with the AGES spectral resolution are 253 km s$^{-1}$ and 205 km s$^{-1}$, respectively.

\begin{deluxetable*}{cccccccc}
\tabletypesize{\scriptsize}
\tablecaption{Observed Properties of Double-peaked AGNs and Offset AGN Candidates in AGES}
\tablewidth{0pt}
\tablecolumns{8}
\tablehead{ 
  \colhead{ID}  & 
  \colhead{Host }   & 
  \colhead{Weighted Mean} &
  \colhead{\oiiiwn/} & 
  \colhead{\niin/} &
  \colhead{\mbox{[\ion{S}{2}] $\lambda\lambda$6716,31}/} &
   \colhead{\mbox{[\ion{O}{1}] $\lambda$6300}/} &
   \colhead{Classification} \\ 
  \colhead{} &
  \colhead{Galaxy} &
   \colhead{Velocity Offset} &
  \colhead{\hb} &
  \colhead{\ha} &
  \colhead{\ha} &
  \colhead{\ha} &
    \colhead{} \\
  \colhead{} &
  \colhead{Redshift$^a$} &
   \colhead{(km s$^{-1}$)} &
  \colhead{} &
    \colhead{} &
      \colhead{} &
      \colhead{} &
  \colhead{}}
\startdata
{\scriptsize NDWFS J143208.27+353255.5} &	0.0834	 &	 $-205.8 \pm \phn 3.3$ &	$7.46 \pm 0.01$	&  $1.31 \pm 0.04$ & $0.44 \pm 0.34$ & $0.12 \pm 1.52$ & Seyfert \\   
		 &  		              &	         $\phn \phd 152.3 \pm \phn 2.8$	  & $6.13 \pm 0.27$	&   $1.46 \pm 0.05$ & $0.52 \pm 0.41$ & $0.05 \pm 2.00$ & Seyfert \\
{\scriptsize NDWFS J143359.71+351020.5} &	0.3372	 &	 $-159.5 \pm 26.8$ & $3.16 \pm 0.23$	&   $0.80 \pm 0.41$ & $1.08 \pm 2.25$ & $0.25 \pm 3.14$ & LINER \\
		 &		              &         $\phn \phn \phd 86.5 \pm 26.1$	  & $2.81 \pm 0.26$ &   $0.80 \pm 0.36$  & $0.58 \pm 1.73$ & $0.15 \pm 3.13$ & Seyfert \\
\hline		 
{\scriptsize NDWFS J143044.06+335224.5} &	0.2297  &	 $\phn \phd 217.7 \pm 23.0$ & $1.86 \pm 0.12$ & 0.75 $\pm$ 0.03  & $0.56 \pm 0.08$ & $0.16 \pm 0.06$ & LINER \\
{\scriptsize NDWFS J143053.69+345836.4} &	0.0839	 &	     $-123.4 \pm \phn 9.8$ & 3.39 $\pm$ 0.10 & 0.86 $\pm$ 0.02 & $0.33 \pm 0.02$ & $0.07 \pm 0.01$ & Seyfert \\ 
{\scriptsize NDWFS J143316.48+353259.3} &	0.1994	 &	    	 $-100.9 \pm \phn 9.1$ &3.10 $\pm$ 0.36	& 0.41 $\pm$ 0.04 & $0.39 \pm 0.26$ & $0.11 \pm 0.19$ & Seyfert \\
{\scriptsize NDWFS J143317.07+344912.0} &	0.3637	 &	    $-143.0 \pm 17.3$ & 4.23 $\pm$ 1.41	& 0.41 $\pm$ 0.03 & $0.86 \pm 0.18$ & $0.07 \pm 0.09$ & ambiguous$^b$ \\
{\scriptsize NDWFS J143710.03+343530.1} &	0.1266	 &	 \phn $-97.6 \pm 11.4$ & 2.41 $\pm$ 0.14	& \phd 0.64 $\pm$ 0.03 & $0.24 \pm 0.02$ & $0.03 \pm 0.01$ & ambiguous
\enddata
\label{tbl:properties}
\tablecomments{The objects are listed in the same order, from top to bottom, as their corresponding spectra in Figure~\ref{fig:offset}.  The first two objects are the double-peaked AGNs, which have two entires each.  The first entry gives the properties of the blueshifted component, while the second gives the properties of the redshifted component.  \\ $^a$The uncertainties on the host galaxy redshifts are 0.0003. \\ $^b$The line ratios, within their errors, are consistent with classification as a Seyfert. }
\end{deluxetable*}

\vspace{.5in}
\subsection{Identification of Five Offset AGN Candidates}
\label{offset}

Next, we searched the AGES spectra for offset AGN candidates.  Velocity offsets in AGN emission lines can be produced not only by offset AGNs, but also by other kinematic effects such as AGN outflows.  An AGN outflow that decelerates with distance from the central AGN will produce a stratified velocity structure, imparting higher velocities to lines with higher ionization potentials, such as \oiiin, and lower velocities to lines with lower ionization potentials, such as \hb and \ha (e.g., \citealt{ZA02.1,KO08.1}).  In contrast, all of the emission lines should exhibit the same velocities if they are produced by the bulk motion of an offset AGN moving within the host galaxy.

We searched the 173 AGN spectra for signatures of offset AGNs that have bulk velocities within their host galaxies.  First, to avoid emission line systems with nonzero velocity differences that are caused by measurement errors, we selected the 14 AGNs with velocity offsets that are different from zero by $> 3\sigma$ for each of the \hbn, \oiiiwn, and \ha emission lines (we note that there are an additional 32 AGNs that have $3\sigma$ velocity offsets in \oiiiw but not \hb or \han; these may be AGN outflows).  Then, we selected the five AGNs that have \hbn, and \oiiiwn, and \ha velocity offsets that are all consistent to within $1\sigma$.  These five AGNs (Figure~\ref{fig:offset}; Table~\ref{tbl:properties}) are our offset AGN candidates.  The inverse-variance-weighted mean velocity offsets of the five offset AGN candidates span $98$ km s$^{-1}$ $< | v_{em} - v_{abs} | < 218$ km s$^{-1}$ (Table~\ref{tbl:properties}).

Although the selection criteria are similar, we note that these velocity-offset narrow AGN emission lines are a different class from the velocity-offset {\it broad} AGN emission lines used in searches for subparsec-scale SMBH binaries and gravitationally recoiling SMBHs (e.g., \citealt{GA84.1,BO07.1,ER12.1,JU13.1,SH13.1}).  Binary and recoiling SMBH searches typically focus on the broad-line region carried with the SMBH at velocities $\simgt 1000$ km s$^{-1}$, while our search for dual SMBHs targets narrow-line velocity offsets, which are up to $\sim$few hundred km s$^{-1}$ for kpc-scale separation dual SMBHs.

\begin{figure}
\begin{center}
\includegraphics[width=9cm]{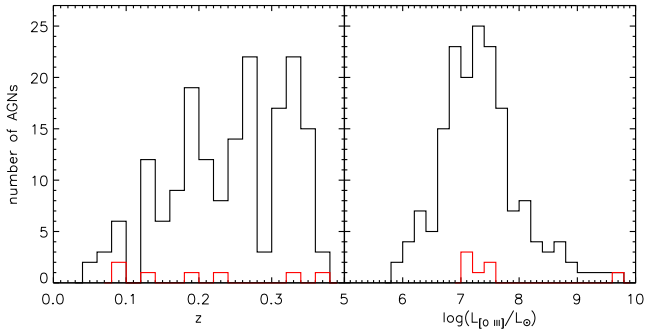}
\end{center}
\caption{Histograms of the redshift (left) and \oiiiw luminosity (right) distributions of the seven offset and dual AGN candidates (red) as compared to the parent population of AGES AGNs (black).  Both populations are consistent with being drawn from the same distribution of redshifts and the same distribution of \oiiiw luminosities.
}
\label{fig:compare}
\end{figure}

\vspace{.5in}
\subsection{Comparison to Parent AGN Population}
\label{parent}

To test whether the velocity offset AGNs have other unique observable characteristics, we compare the subset of seven offset and dual AGN candidates to the parent population of 173 Type 2 AGNs in AGES.  The redshift distributions of the two samples, shown in Figure~\ref{fig:compare}, are consistent.  According to a Kolmogorov-Smirnov test, there is a 66$\%$ probability that the redshifts of the seven dual SMBH candidates were drawn from the same distribution as the redshifts of the AGES AGN parent sample.
We also examine the \oiiiw luminosities (measured in Section~\ref{agnmeasure}) and find a probability of 66$\%$ that the \oiiiw luminosities of the dual SMBH candidates and AGN parent population were drawn from the same distribution.  
Finally, we note that the spectral resolutions are too low to measure resolved gas velocity dispersions for the entire sample.

In general, we find that the dual SMBH candidate subpopulation has redshifts and \oiiiw luminosities that are similar to those of the parent AGN population.  However, it is striking that the same offset AGN candidate, NDWFS J143317.07+344912.0, has both the second highest redshift ($z=0.3637$) and the largest \oiiiw luminosity ($L_{\mathrm{[O \, III]}} =5.2 \times 10^{9} \, L_\odot$) of the entire sample of 173 Type 2 AGNs in AGES.  We discuss additional interesting features of this system in the following section.

\begin{figure}
\begin{center}
\includegraphics[width=9cm]{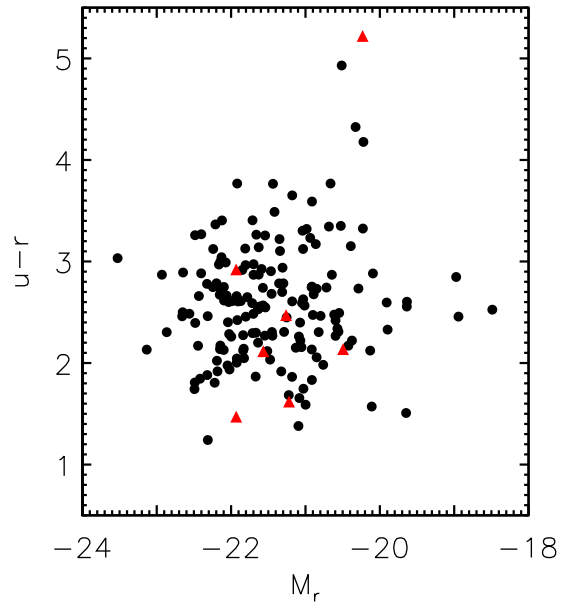}
\end{center}
\caption{Color-magnitude diagram of $u-r$ color and absolute $r$-band magnitude M$_r$, based on SDSS photometry of the host galaxies of all 173 AGES AGNs in our sample.  The seven offset and dual AGN candidates (red triangles) have host galaxies with similar colors and absolute magnitudes to those of the overall AGN population (black circles).  
}
\label{fig:colormag}
\end{figure}

\subsection{Host Galaxies of Offset and Dual AGN Candidates}

We also use SDSS photometry to examine the host galaxies of the offset and dual AGN candidates, and we find that the candidates' host galaxies colors and absolute magnitudes are similar to those of the parent population of AGN host galaxies (Figure~\ref{fig:colormag}).  For $u-r$ colors (M$_r$ absolute magnitudes), there is a 64$\%$ (72$\%$) probability that the candidates were drawn from the same distribution as all AGN host galaxies.  One interesting exception is that the reddest AGN host galaxy ($u-r=5.22$) is the host galaxy of one of the double-peaked AGNs, NDWFS J143359.71+351020.5.  If this galaxy is shown to indeed host dual AGNs, it will be interesting to examine whether the galaxy is dust-reddened or whether there is a lack of ongoing star formation to reconcile with the fueling of two AGNs.

\begin{figure}
\begin{center}
\includegraphics[width=9cm]{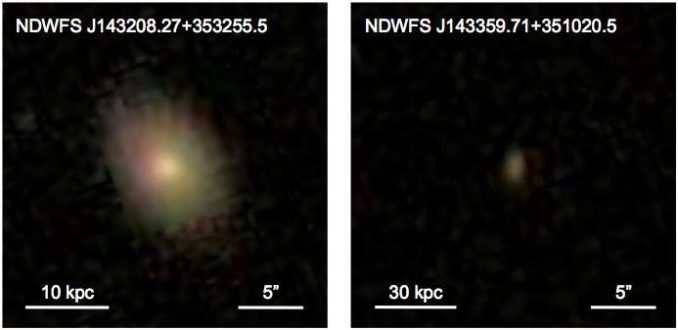}
\end{center}
\caption{SDSS {\it gri} color-composite images of the two double-peaked AGN host galaxies in AGES.  Each panel is $25^{\prime\prime} \times 25^{\prime\prime}$, with North up and East to the left.  The scale bars in the lower left hand corners illustrate the physical distance scales at the redshifts of the host galaxies.
}
\label{fig:dual_image}
\end{figure}

\begin{figure}
\begin{center}
\includegraphics[width=9cm]{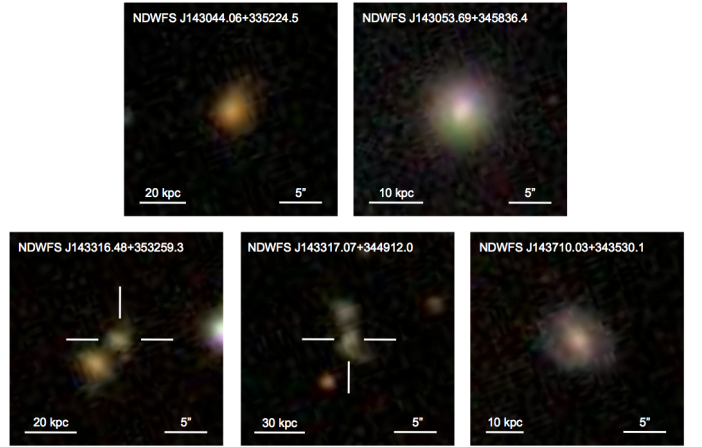}
\end{center}
\caption{Same as Figure~\ref{fig:dual_image}, but for the five offset AGN candidates in AGES.  In the systems that have multiple interacting galaxies, the primary AGES sources are marked with crosshairs.}
\label{fig:offset_image}
\end{figure}

Further, we visually inspect the SDSS {\it gri} color-composite images of the host galaxies for evidence of galaxy mergers that may have caused the AGN velocity offsets we measure in the seven dual SMBH candidates.  The images of the two double-peaked AGNs are shown in Figure~\ref{fig:dual_image}, and the images of the five offset AGN candidates are shown in Figure~\ref{fig:offset_image}.  Whereas only $9^{+3}_{-2}\%$ (16/173) of the AGES AGN host galaxies have a companion within $5^{\prime\prime}$, this figure increases to $29^{+26}_{-19}\%$ (2/7) for the dual SMBH candidates.

The two offset AGN candidates with nearby companions are NDWFS J143316.48+353259.3 and NDWFS J143317.07+344912.0, which is also the galaxy with the second highest redshift and the largest \oiiiw luminosity in the AGN sample (Section~\ref{parent}). Notably, neither companion is so close that it overlaps with the $1\farcs5$ diameter Hectospec fiber used to obtain the AGES spectrum of the primary galaxy.  Therefore, the stellar continuum and the AGN emission features we measure in these AGES spectra arise solely from the primary galaxy and not from, e.g., a combination of continuum from the primary and AGN emission from the companion.  We discuss each companion in detail in Sections~\ref{companion1} and \ref{companion2}.

We find evidence that dual SMBH candidate host galaxies have a $\sim3$ times higher probability of having companions, as compared to the general population of AGN host galaxies.  This greater probability suggests that mergers may be at the root of the AGN velocity offsets we measure here. We note that in the cases of the two offset AGN candidates that have companions, both the primary and companion galaxies have faint, spatially-extended features that resemble tidal tails.  This implies that both systems may be galaxy mergers in progress that have already had their first pericenter passages.  The initial collision or collisions could have produced the velocity offsets we measure in the AGES spectrum by disrupting the velocity of the stars relative to the central AGN (e.g., \citealt{HI12.1}), by producing tidal features that skew the observed velocity of the stars due to projection effects, or by jostling the AGN to a different velocity than the stars.

\subsubsection{Companion to Offset AGN Candidate NDWFS J143316.48+353259.3}
\label{companion1}

The southeast companion to NDWFS J143316.48+353259.3 is separated by 15 kpc and is 1.5 times more luminous than the primary.
The companion has an AGES spectrum, and we measure this companion's redshift and its emission line fluxes and velocity offsets as in Section 2.  We note that the companion, the galaxy NDWFS J143316.70+353256.4, did not survive our initial quality cut (Section~\ref{quality}). Although the emission lines are detected at $>2\sigma$ significance, the \ha equivalent width of 2.3 \AA \, is below our quality cutoff of 5 \AA.

\begin{figure*}
\begin{center}
\subfigure{\includegraphics[height=4.1cm]{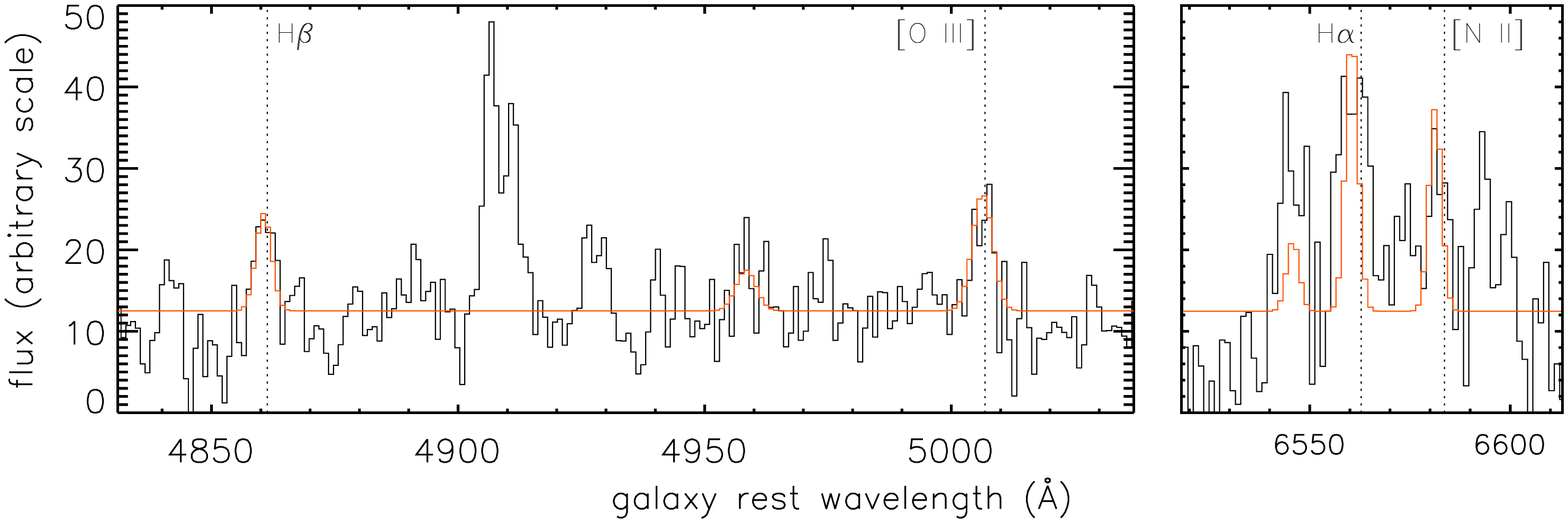}}
\hspace{-0.2cm}
\subfigure{\includegraphics[height=4.1cm]{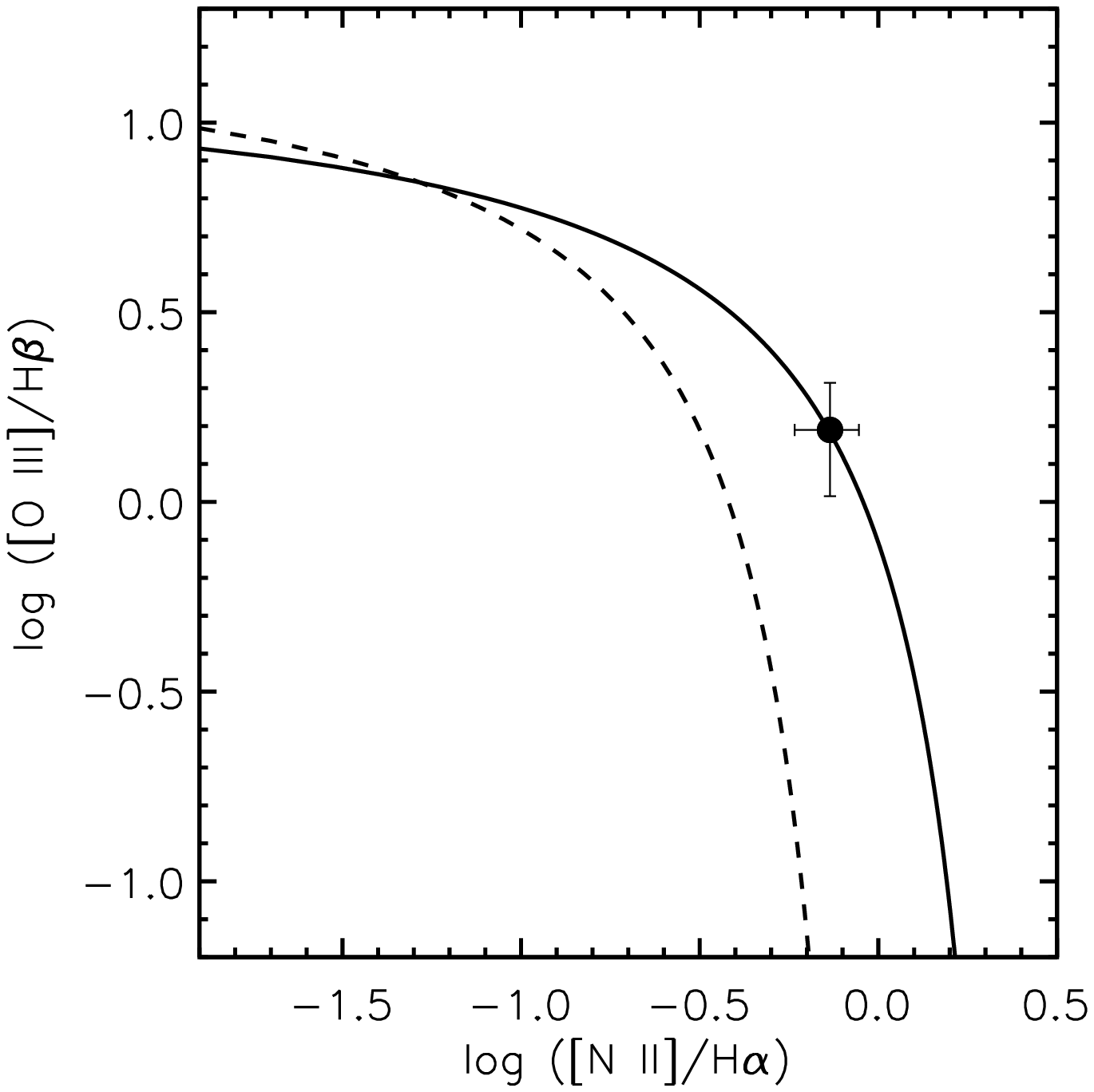}} 
\end{center}
\caption{\footnotesize{As Figure~\ref{fig:offset} (left), but showing the spectrum and fits for NDWFS J143316.70+353256.4, the companion galaxy to the offset AGN candidate NDWFS J143316.48+353259.3. Both galaxies are at nearly the same redshift, and the companion's emission lines have a line-of-sight velocity offset of $-35.6 \pm 18.9$ km s$^{-1}$.  Hence, the companion may host an offset AGN in addition to the offset AGN candidate in the primary galaxy.  As Figure~\ref{fig:BPT} (right), the BPT diagram  shows that the companion galaxy's emission line flux ratios (solid point) are in the AGN regime just above the theoretical maximum for starbursts (solid line; \citealt{KE01.2}), but the error bars allow the possibility that the galaxy is an AGN -- starburst composite (region between solid line and dashed line; \citealt{KA03.1}). }}
\label{fig:companion}
\end{figure*}

We find that the companion displays an almost identical redshift to the primary (Table~\ref{tbl:companions}), placing it at a velocity $-0.4 \pm 94.4$ km s$^{-1}$ (blueward) of the primary galaxy.  Based on its flux ratios of $\oiiiwn/\hb=1.55 \pm 0.51$ and $\niin/\ha=0.73 \pm 0.15$, the companion just makes the classification as a pure AGN \citep{KE01.2}, but the errors on the line flux measurements include the possibility of an AGN -- star formation composite system (\citealt{KA03.1,KE06.1}; Figure~\ref{fig:companion}).  

The emission lines have line-of-sight velocity offsets ($\Delta v=v_{em}-v_{abs}$) of $\Delta v_{\hbn}=-34.5 \pm 33.0$ km s$^{-1}$, $\Delta v_{\oiiiwn}=-22.4 \pm 29.4$ km s$^{-1}$, and $\Delta v_{\han}=-58.0 \pm 37.2$ km s$^{-1}$.  The inverse-variance-weighted mean of the velocity offsets is $\Delta v=-35.6 \pm 18.9$ km s$^{-1}$.  The velocity offsets of the three emission lines are all consistent to within $1\sigma$, suggestive of the bulk motion of an offset AGN (see Section~\ref{offset}), but each velocity offset measurement is also $<2\sigma$ from zero.  Either the AGN has a small line-of-sight velocity offset relative to systemic or it has no offset; additional observations with improved spectral resolution could distinguish between these scenarios.

The similar emission-line velocity offsets in the two galaxies ($-100.9 \pm 9.1$ km s$^{-1}$ for the primary and $-35.6 \pm 18.9$ km s$^{-1}$ for the companion) and the nearly identical redshifts derived from the two sets of stellar absorption features (the difference is $0.4 \pm 94.4$ km s$^{-1}$) suggests that the two galaxies are dynamically linked.  When combined with the tidal features visible in the SDSS image, this is evidence that the system is one or more pericenter passages into the merger process.

\subsubsection{Companion to Offset AGN Candidate NDWFS J143317.07+344912.0}
\label{companion2}

Located 16 kpc from NDWFS J143317.07+344912.0 is a northern companion that is 1.7 times less luminous than the primary.
Although there are not AGES or SDSS spectra for the companion, the photometric redshifts from SDSS suggest that the companion is at roughly the same redshift as the primary (Table~\ref{tbl:companions}; photometric redshifts for NDWFS J143316.48+353259.3 are also shown for comparison, since its companion is spectroscopically confirmed).  However, a spectroscopic redshift is required to 
determine whether the companion is indeed associated with NDWFS J143317.07+344912.0.

\begin{deluxetable*}{cccccc}
\tabletypesize{\scriptsize}
\tablecaption{Properties of Offset AGN Candidates with Companion Galaxies}
\tablewidth{0pt}
\tablecolumns{4}
\tablehead{ 
  \colhead{ID}  & 
  \colhead{$z_\mathrm{spectroscopic}$}   & 
  \colhead{$z_\mathrm{photo\_template}$} & 
  \colhead{$z_\mathrm{photo\_neural}$}  &
    \colhead{Separation (kpc)}  &
    \colhead{$L_r (10^{10} L_\odot)$}  }
 \startdata
{\scriptsize NDWFS J143316.48+353259.3} &	0.199419	$\pm$ 0.0003 & $0.185 \pm 0.058$ & $0.220 \pm 0.079$  & 15 & 1.3 \\
{\scriptsize NDWFS J143316.70+353256.4$^a$} & 0.199418 $\pm$ 0.0003 & $0.152 \pm 0.026$ & $0.162 \pm 0.045$ & & 1.9 \\
\hline
{\scriptsize NDWFS J143317.07+344912.0} &	0.3637	$\pm$ 0.0003 & $0.302 \pm 0.037$ & $0.252 \pm 0.085$  & 16 & 4.7 \\
{\scriptsize northern companion} & N/A & $0.267 \pm 0.040$ & \phd $0.305 \pm 0.101$  & & 2.8
\enddata
\label{tbl:companions}
\tablecomments{The spectroscopic redshifts $z_\mathrm{spectroscopic}$ are measured from AGES spectra, while the SDSS photometric redshifts are measured with the template fitting method ($z_\mathrm{photo\_template}$) and with a Neural Network method ($z_\mathrm{photo\_neural}$). We use the spectroscopic redshifts in measurements of the separations between the two galaxies and the $r$-band luminosities $L_r$, except in the case of the northern companion to NDWFS J143317.07+344912.0.  There we use the weighted mean of the photometric redshifts, $z=0.272$.
\\ $^a$This galaxy is the southeast companion to the offset AGN candidate above, NDWFS J143316.48+353259.3.}
\end{deluxetable*}

\subsection{Comparison to Other Samples of Offset and Dual AGN Candidates}
\label{compare}

We find two double-peaked AGNs out of the 173 Type 2 AGNs at $z<0.37$ in AGES, which corresponds to a rate of $1.2^{+1.5}_{-0.8}\%$.  For comparison, the fraction of double-peaked AGNs is $2.2^{+2.8}_{-1.4}\%$ (2/91) in the DEEP2 sample of Type 2 AGNs in red galaxies at $0.34 < z < 0.82$ \citep{CO09.1}, $1.3^{+0.1}_{-0.1}\%$ (87/6780) in SDSS Type 2 AGNs at $z \leq 0.15$ \citep{WA09.1}, and $1.1^{+0.09}_{-0.09}\%$ (167/14756) in SDSS Type 2 AGNs and Type 2 quasars at $z <0.83$ \citep{LI10.1}.  Since all of the double-peaked AGNs in AGES, DEEP2, and SDSS have velocity separations $>150$ km s$^{-1}$, the differences in the surveys' spectral resolutions ($R\sim1000$ for AGES, $R\sim5000$ for DEEP2, and $R\sim1800$ for SDSS) do not bias the comparison of the fractions of double-peaked AGNs found in each.  

The three surveys do not probe the same spatial scales (SDSS uses $3^{\prime\prime}$ fibers, which corresponds to 5.5 kpc at typical redshift $z=0.1$; AGES uses $1\farcs5$ fibers, which corresponds to 5.0 kpc at typical redshift $z=0.2$; and DEEP2 uses slits with 1$^{\prime\prime}$ widths and 7$^{\prime\prime}$ average lengths, which correspond to 7 kpc and 50 kpc, respectively, at typical redshift $z=0.6$ of the double-peaked and offset sample; \citealt{NE12.1}), allowing the possibility that the double peaks in each survey are produced by effects on different spatial scales.  Although DEEP2 can probe scales of 50 kpc, the sources of the double-peaked emission exist on $\sim1$ kpc scales. The two double-peaked AGNs in DEEP2 correspond to double emission components with spatial separations of 1.2 kpc and 2.3 kpc \citep{CO09.1}, and long-slit observations of 81 double-peaked AGNs in SDSS showed that the double emission components have a typical separation of 1.1 kpc (with a range of 0.2 kpc to 5.5 kpc, where the maximum is 15 kpc for a $3^{\prime\prime}$ SDSS fiber; \citealt{CO12.1}).  As a result, the different spatial scales explored by SDSS, AGES, and DEEP2 should not significantly bias the relative numbers of offset and dual AGN candidates found in each. 

Finally, the DEEP2 search was limited to red galaxies. If we match the AGES AGN sample to the color ($-0.6 < u-r < 5.8$) and absolute magnitude ($-27.4 < \mathrm{M}_r < -21.7$) range of the DEEP2 sample, we find zero double-peaked system out of 70 AGNs.

We also find five offset AGN candidates, which is $2.9^{+1.9}_{-0.8}\%$ of our sample of 173 Type 2 AGNs at $z<0.37$.  For comparison, the fraction is $33^{+5}_{-5}\%$ (30/91) for offset AGN candidates in the DEEP2 sample of Type 2 AGNs in red galaxies at $0.34 < z < 0.82$ \citep{CO09.1}.  A major source of this discrepancy is that the DEEP2 spectra have five times the spectral resolution ($R\sim5000$) of the AGES spectra; the DEEP2 offset AGN candidates have velocity offsets down to 37 km s$^{-1}$ with $3\sigma$ significance, whereas the AGES offset AGN candidates have velocity offsets down to 98 km s$^{-1}$ with $3\sigma$ significance.  A more accurate comparison is the $15^{+5}_{-3}\%$ (14/91) fraction of DEEP2 Type 2 AGNs that have velocity offsets $>98$ km s$^{-1}$ to the $2.9^{+3.5}_{-1.0}\%$ (2/70) fraction of offset AGN candidates in an AGES AGN sample that is matched in color and absolute magnitude (as described in the previous paragraph) to the DEEP2 sample.  

When the AGES and DEEP2 samples are matched in color, absolute magnitude, and minimum velocity offset, there is clear evidence for redshift evolution in the fractions of offset and dual AGN candidates.  For the subsamples matched in this way, the fraction of offset (dual) AGN candidates increases with redshift, from $2.9\%$ ($0\%$) in AGES  at $z<0.37$ and mean redshift $\bar{z}=0.25$, to $15\%$ ($2.2\%$) in DEEP2 at $0.34<z<0.82$ and mean redshift $\bar{z}=0.7$.  When the offset and dual AGN candidates are combined, the overall fractions increase by a factor of $\sim6$ from $2.9^{+3.6}_{-1.9}\%$ (2/70) in AGES to $18^{+5}_{-5}\%$ (16/91) in DEEP2.
This trend is similar to the increase in galaxy merger fraction (e.g., \citealt{CO03.2,LI08.1,LO11.1}) with increasing redshift over this redshift range, as expected if these candidates are indeed offset and dual AGNs.

If the five AGES candidates are offset AGNs, then we expect there to be equal numbers of redshifted and blueshifted AGNs, as was the case for the offset AGN candidates in DEEP2 \citep{CO09.1}.  In AGES, $20^{+32}_{-17}\%$ (1/5) of the candidates exhibit redshifted velocity offsets and $80^{+17}_{-32}\%$ (4/5) of the candidates exhibit blueshifted velocity offsets.  Given the small sample size, this could be roughly consistent with an even distribution, or it could be an indication of AGN outflows biasing the sample.  AGN outflows are known to produce an overabundance of observed blueshifted emission lines, since redshifted lines are often obscured by the AGN torus (e.g., \citealt{ZA02.1}).

\section{Conclusions}

We have searched 8136 AGES galaxy spectra at $z<0.37$ for spectroscopic signatures of dual SMBHs.  Dual SMBHs that are accreting gas as dual AGNs can produce double-peaked AGN emission lines, while an offset AGN, which is a dual-SMBH system where only one SMBH is active, can produce AGN emission lines with bulk line-of-sight velocity offsets relative to the host galaxy stellar absorption features.  Out of 173 Type 2 AGN spectra at $z<0.37$, we find two double-peaked AGNs and five offset AGN candidates. The two double-peaked AGNs have line-of-sight velocity separations between their double peaks of 246 km s$^{-1}$ and 358 km s$^{-1}$, while the five offset AGN candidates have velocity offsets of $98$ km s$^{-1}$ $< | v_{em} - v_{abs} | < 218$ km s$^{-1}$.  

We find that $1.2^{+1.5}_{-0.8}\%$ of the AGES AGNs are double-peaked AGNs and $2.9^{+1.9}_{-0.8}\%$ are offset AGN candidates.  When the AGES and DEEP2 samples are matched in color, absolute magnitude, and minimum velocity offset, the dual supermassive black hole candidate fraction increases by a factor of $\sim6$ (from 2/70 to 16/91, or $2.9^{+3.6}_{-1.9}\%$ to $18^{+5}_{-5}\%$) from the AGES sample at a mean redshift of $\bar{z}=0.25$ to the DEEP2 sample at $\bar{z}=0.7$.  This trend could be understood if velocity-offset narrow AGN emission lines are associated with galaxy mergers, since the galaxy merger fraction also increases with redshift over this range.

In an additional sign of a link between velocity offsets and galaxy mergers, we find tentative evidence that the offset and dual AGN candidates are $\sim3$ times more likely (2/7, or $29^{+26}_{-19}\%$) to be hosted by galaxies with a companion within $5^{\prime\prime}$ than the parent AGES AGN population (16/173, or $9^{+3}_{-2}\%$). Since our sample consists of only seven offset and dual AGN candidates, a larger sample is needed to confirm this result. Further, we find that two of the offset AGN candidates have companions, and one of the companions may host an offset AGN itself that has a similar velocity offset to the primary galaxy's AGN.  Both sets of primary and companion galaxies also have morphological signs of disturbance, suggesting that they have already interacted.  In the dual SMBH scenario, such an interaction could perturb the central AGN to a different velocity than systemic.

While the seven objects we find are compelling candidates for offset and dual AGNs, their line profiles could also be produced by kinematics in the narrow-line region such as outflows and rotating disks (e.g., \citealt{VE01.1,CR10.1}).  Follow-up studies of the SDSS double-peaked AGNs have showcased the utility of long-slit spectroscopy \citep{LI10.2,GR11.1,SH11.1,CO12.1}, integral-field unit spectroscopy \citep{FU12.1}, near-infrared imaging \citep{LI10.2,FU11.1,MC11.1,RO11.1,SH11.1,FU12.1}, radio observations \citep{FU11.3,TI11.1}, and X-ray observations \citep{CO11.2,LI13.1} in distinguishing narrow-line region kinematics from bona fide dual AGNs.  While radio and X-ray observations are particularly promising avenues for confirmations of dual AGNs, the {\it Chandra} observations to date have underscored the difficulties in identifying dual AGNs when the AGNs are closely separated and faint or have high X-ray obscurations \citep{CO11.2,LI13.1}.  Building on the strategies honed for SDSS double-peaked AGNs, careful follow-up observations of the AGES offset and dual AGN candidates presented here would pinpoint which systems are dual SMBHs, which would help clarify our understanding of spectroscopic signatures of dual SMBHs and advance these systems as new probes of galaxy evolution.

\acknowledgements{J.M.C. was supported by an NSF Astronomy and Astrophysics Postdoctoral Fellowship under award AST-1102525. J.E.G. is partially supported by an Alfred P. Sloan Fellowship.  Observations reported here were obtained at the MMT Observatory, a joint facility of the University of Arizona and the Smithsonian Institution.} 

{\it Facility:} \facility{MMT (Hectospec)}

\bibliographystyle{apj}

\begin{thebibliography}{62}
\expandafter\ifx\csname natexlab\endcsname\relax\def\natexlab#1{#1}\fi

\bibitem[{{Alonso-Herrero} {et~al.}(2000){Alonso-Herrero}, {Rieke}, {Rieke}, \&
  {Shields}}]{AL00.1}
{Alonso-Herrero}, A., {Rieke}, M.~J., {Rieke}, G.~H., \& {Shields}, J.~C. 2000,
  \apj, 530, 688

\bibitem[{{Baldwin} {et~al.}(1981){Baldwin}, {Phillips}, \&
  {Terlevich}}]{BA81.1}
{Baldwin}, J.~A., {Phillips}, M.~M., \& {Terlevich}, R. 1981, \pasp, 93, 5

\bibitem[{{Barrows} {et~al.}(2013){Barrows}, {Sandberg Lacy}, {Kennefick},
  {Comerford}, {Kennefick}, \& {Berrier}}]{BA13.1}
{Barrows}, R.~S., {Sandberg Lacy}, C.~H., {Kennefick}, J., {Comerford}, J.~M.,
  {Kennefick}, D., \& {Berrier}, J.~C. 2013, \apj, 769, 95

\bibitem[{{Barrows} {et~al.}(2012){Barrows}, {Stern}, {Madsen}, {Harrison},
  {Assef}, {Comerford}, {Cushing}, {Fassnacht}, {Gonzalez}, {Griffith},
  {Hickox}, {Kirkpatrick}, \& {Lagattuta}}]{BA12.1}
{Barrows}, R.~S., {Stern}, D., {Madsen}, K., {Harrison}, F., {Assef}, R.~J.,
  {Comerford}, J.~M., {Cushing}, M.~C., {Fassnacht}, C.~D., {Gonzalez}, A.~H.,
  {Griffith}, R., {Hickox}, R., {Kirkpatrick}, J.~D., \& {Lagattuta}, D.~J.
  2012, \apj, 744, 7

\bibitem[{{Barth} {et~al.}(2008){Barth}, {Bentz}, {Greene}, \& {Ho}}]{BA08.1}
{Barth}, A.~J., {Bentz}, M.~C., {Greene}, J.~E., \& {Ho}, L.~C. 2008, \apjl,
  683, L119

\bibitem[{{Begelman} {et~al.}(1980){Begelman}, {Blandford}, \& {Rees}}]{BE80.1}
{Begelman}, M.~C., {Blandford}, R.~D., \& {Rees}, M.~J. 1980, \nat, 287, 307

\bibitem[{{Blecha} {et~al.}(2013){Blecha}, {Loeb}, \& {Narayan}}]{BL13.1}
{Blecha}, L., {Loeb}, A., \& {Narayan}, R. 2013, \mnras, 429, 2594

\bibitem[{{Bonning} {et~al.}(2007){Bonning}, {Shields}, \&
  {Salviander}}]{BO07.1}
{Bonning}, E.~W., {Shields}, G.~A., \& {Salviander}, S. 2007, \apjl, 666, L13

\bibitem[{{Boroson}(2002)}]{BO02.1}
{Boroson}, T.~A. 2002, \apj, 565, 78

\bibitem[{{Comerford} {et~al.}(2009{\natexlab{a}}){Comerford}, {Gerke},
  {Newman}, {Davis}, {Yan}, {Cooper}, {Faber}, {Koo}, {Coil}, {Rosario}, \&
  {Dutton}}]{CO09.1}
{Comerford}, J.~M., {Gerke}, B.~F., {Newman}, J.~A., {Davis}, M., {Yan}, R.,
  {Cooper}, M.~C., {Faber}, S.~M., {Koo}, D.~C., {Coil}, A.~L., {Rosario},
  D.~J., \& {Dutton}, A.~A. 2009{\natexlab{a}}, \apj, 698, 956

\bibitem[{{Comerford} {et~al.}(2012){Comerford}, {Gerke}, {Stern}, {Cooper},
  {Weiner}, {Newman}, {Madsen}, \& {Barrows}}]{CO12.1}
{Comerford}, J.~M., {Gerke}, B.~F., {Stern}, D., {Cooper}, M.~C., {Weiner},
  B.~J., {Newman}, J.~A., {Madsen}, K., \& {Barrows}, R.~S. 2012, \apj, 753, 42

\bibitem[{{Comerford} {et~al.}(2009{\natexlab{b}}){Comerford}, {Griffith},
  {Gerke}, {Cooper}, {Newman}, {Davis}, \& {Stern}}]{CO09.3}
{Comerford}, J.~M., {Griffith}, R.~L., {Gerke}, B.~F., {Cooper}, M.~C.,
  {Newman}, J.~A., {Davis}, M., \& {Stern}, D. 2009{\natexlab{b}}, \apjl, 702,
  L82

\bibitem[{{Comerford} {et~al.}(2011){Comerford}, {Pooley}, {Gerke}, \&
  {Madejski}}]{CO11.2}
{Comerford}, J.~M., {Pooley}, D., {Gerke}, B.~F., \& {Madejski}, G.~M. 2011,
  \apjl, 737, L19+

\bibitem[{{Conselice} {et~al.}(2003){Conselice}, {Bershady}, {Dickinson}, \&
  {Papovich}}]{CO03.2}
{Conselice}, C.~J., {Bershady}, M.~A., {Dickinson}, M., \& {Papovich}, C. 2003,
  \aj, 126, 1183

\bibitem[{{Cool} {et~al.}(2012){Cool}, {Eisenstein}, {Kochanek}, {Brown},
  {Caldwell}, {Dey}, {Forman}, {Hickox}, {Jannuzi}, {Jones}, {Moustakas}, \&
  {Murray}}]{CO12.2}
{Cool}, R.~J., {Eisenstein}, D.~J., {Kochanek}, C.~S., {Brown}, M.~J.~I.,
  {Caldwell}, N., {Dey}, A., {Forman}, W.~R., {Hickox}, R.~C., {Jannuzi},
  B.~T., {Jones}, C., {Moustakas}, J., \& {Murray}, S.~S. 2012, \apj, 748, 10

\bibitem[{{Crenshaw} {et~al.}(2010){Crenshaw}, {Schmitt}, {Kraemer},
  {Mushotzky}, \& {Dunn}}]{CR10.1}
{Crenshaw}, D.~M., {Schmitt}, H.~R., {Kraemer}, S.~B., {Mushotzky}, R.~F., \&
  {Dunn}, J.~P. 2010, \apj, 708, 419

\bibitem[{{Dopita} \& {Sutherland}(1995)}]{DO95.1}
{Dopita}, M.~A., \& {Sutherland}, R.~S. 1995, \apj, 455, 468

\bibitem[{{Eracleous} {et~al.}(2012){Eracleous}, {Boroson}, {Halpern}, \&
  {Liu}}]{ER12.1}
{Eracleous}, M., {Boroson}, T.~A., {Halpern}, J.~P., \& {Liu}, J. 2012, \apjs,
  201, 23

\bibitem[{{Fabbiano} {et~al.}(2011){Fabbiano}, {Wang}, {Elvis}, \&
  {Risaliti}}]{FA11.1}
{Fabbiano}, G., {Wang}, J., {Elvis}, M., \& {Risaliti}, G. 2011, \nat, 477, 431

\bibitem[{{Filippenko} \& {Halpern}(1984)}]{FI84.1}
{Filippenko}, A.~V., \& {Halpern}, J.~P. 1984, \apj, 285, 458

\bibitem[{{Fu} {et~al.}(2011{\natexlab{a}}){Fu}, {Myers}, {Djorgovski}, \&
  {Yan}}]{FU11.1}
{Fu}, H., {Myers}, A.~D., {Djorgovski}, S.~G., \& {Yan}, L. 2011{\natexlab{a}},
  \apj, 733, 103

\bibitem[{{Fu} {et~al.}(2012){Fu}, {Yan}, {Myers}, {Stockton}, {Djorgovski},
  {Aldering}, \& {Rich}}]{FU12.1}
{Fu}, H., {Yan}, L., {Myers}, A.~D., {Stockton}, A., {Djorgovski}, S.~G.,
  {Aldering}, G., \& {Rich}, J.~A. 2012, \apj, 745, 67

\bibitem[{{Fu} {et~al.}(2011{\natexlab{b}}){Fu}, {Zhang}, {Assef}, {Stockton},
  {Myers}, {Yan}, {Djorgovski}, {Wrobel}, \& {Riechers}}]{FU11.3}
{Fu}, H., {Zhang}, Z.-Y., {Assef}, R.~J., {Stockton}, A., {Myers}, A.~D.,
  {Yan}, L., {Djorgovski}, S.~G., {Wrobel}, J.~M., \& {Riechers}, D.~A.
  2011{\natexlab{b}}, \apjl, 740, L44+

\bibitem[{{Gaskell}(1984)}]{GA84.1}
{Gaskell}, C.~M. 1984, New York Academy Sciences Annals, 422, 349

\bibitem[{{Gerke} {et~al.}(2007){Gerke}, {Newman}, {Lotz}, {Yan}, {Barmby},
  {Coil}, {Conselice}, {Ivison}, {Lin}, {Koo}, {Nandra}, {Salim}, {Small},
  {Weiner}, {Cooper}, {Davis}, {Faber}, \& {Guhathakurta}}]{GE07.1}
{Gerke}, B.~F., {Newman}, J.~A., {Lotz}, J., {Yan}, R., {Barmby}, P., {Coil},
  A.~L., {Conselice}, C.~J., {Ivison}, R.~J., {Lin}, L., {Koo}, D.~C.,
  {Nandra}, K., {Salim}, S., {Small}, T., {Weiner}, B.~J., {Cooper}, M.~C.,
  {Davis}, M., {Faber}, S.~M., \& {Guhathakurta}, P. 2007, \apjl, 660, L23

\bibitem[{{Greene} \& {Ho}(2005)}]{GR05.1}
{Greene}, J.~E., \& {Ho}, L.~C. 2005, \apj, 627, 721

\bibitem[{{Greene} {et~al.}(2011){Greene}, {Zakamska}, {Ho}, \&
  {Barth}}]{GR11.1}
{Greene}, J.~E., {Zakamska}, N.~L., {Ho}, L.~C., \& {Barth}, A.~J. 2011, \apj,
  732, 9

\bibitem[{{Hao} {et~al.}(2005){Hao}, {Strauss}, {Tremonti}, {Schlegel},
  {Heckman}, {Kauffmann}, {Blanton}, {Fan}, {Gunn}, {Hall}, {Ivezi{\'c}},
  {Knapp}, {Krolik}, {Lupton}, {Richards}, {Schneider}, {Strateva}, {Zakamska},
  {Brinkmann}, {Brunner}, \& {Szokoly}}]{HA05.1}
{Hao}, L., {Strauss}, M.~A., {Tremonti}, C.~A., {Schlegel}, D.~J., {Heckman},
  T.~M., {Kauffmann}, G., {Blanton}, M.~R., {Fan}, X., {Gunn}, J.~E., {Hall},
  P.~B., {Ivezi{\'c}}, {\v Z}., {Knapp}, G.~R., {Krolik}, J.~H., {Lupton},
  R.~H., {Richards}, G.~T., {Schneider}, D.~P., {Strateva}, I.~V., {Zakamska},
  N.~L., {Brinkmann}, J., {Brunner}, R.~J., \& {Szokoly}, G.~P. 2005, \aj, 129,
  1783

\bibitem[{{Heckman}(1980)}]{HE80.1}
{Heckman}, T.~M. 1980, \aap, 87, 152

\bibitem[{{Hiner} {et~al.}(2012){Hiner}, {Canalizo}, {Wold}, {Brotherton}, \&
  {Cales}}]{HI12.1}
{Hiner}, K.~D., {Canalizo}, G., {Wold}, M., {Brotherton}, M.~S., \& {Cales},
  S.~L. 2012, \apj, 756, 162

\bibitem[{{Ho}(2008)}]{HO08.2}
{Ho}, L.~C. 2008, \araa, 46, 475

\bibitem[{{Ho} {et~al.}(1997){Ho}, {Filippenko}, {Sargent}, \& {Peng}}]{HO97.1}
{Ho}, L.~C., {Filippenko}, A.~V., {Sargent}, W.~L.~W., \& {Peng}, C.~Y. 1997,
  \apjs, 112, 391

\bibitem[{{Jannuzi} \& {Dey}(1999)}]{JA99.2}
{Jannuzi}, B.~T., \& {Dey}, A. 1999, in ASP Conf. Ser. 191, Photometric
  Redshifts and the Detection of High Redshift Galaxies, ed R. Weymann, L.
  Storrie-Lombardi, M. Sawicki, and R. Brunner (San Francisco, CA: ASP), 111

\bibitem[{{Ju} {et~al.}(2013){Ju}, {Greene}, {Rafikov}, {Bickerton}, \&
  {Badenes}}]{JU13.1}
{Ju}, W., {Greene}, J.~E., {Rafikov}, R.~R., {Bickerton}, S.~J., \& {Badenes},
  C. 2013, ArXiv e-prints

\bibitem[{{Kauffmann} {et~al.}(2003){Kauffmann}, {Heckman}, {Tremonti},
  {Brinchmann}, {Charlot}, {White}, {Ridgway}, {Brinkmann}, {Fukugita}, {Hall},
  {Ivezi{\'c}}, {Richards}, \& {Schneider}}]{KA03.1}
{Kauffmann}, G., {Heckman}, T.~M., {Tremonti}, C., {Brinchmann}, J., {Charlot},
  S., {White}, S.~D.~M., {Ridgway}, S.~E., {Brinkmann}, J., {Fukugita}, M.,
  {Hall}, P.~B., {Ivezi{\'c}}, {\v Z}., {Richards}, G.~T., \& {Schneider},
  D.~P. 2003, \mnras, 346, 1055

\bibitem[{{Kewley} {et~al.}(2001){Kewley}, {Dopita}, {Sutherland}, {Heisler},
  \& {Trevena}}]{KE01.2}
{Kewley}, L.~J., {Dopita}, M.~A., {Sutherland}, R.~S., {Heisler}, C.~A., \&
  {Trevena}, J. 2001, \apj, 556, 121

\bibitem[{{Kewley} {et~al.}(2006){Kewley}, {Groves}, {Kauffmann}, \&
  {Heckman}}]{KE06.1}
{Kewley}, L.~J., {Groves}, B., {Kauffmann}, G., \& {Heckman}, T. 2006, \mnras,
  372, 961

\bibitem[{{Kochanek} {et~al.}(2012){Kochanek}, {Eisenstein}, {Cool},
  {Caldwell}, {Assef}, {Jannuzi}, {Jones}, {Murray}, {Forman}, {Dey}, {Brown},
  {Eisenhardt}, {Gonzalez}, {Green}, \& {Stern}}]{KO12.3}
{Kochanek}, C.~S., {Eisenstein}, D.~J., {Cool}, R.~J., {Caldwell}, N., {Assef},
  R.~J., {Jannuzi}, B.~T., {Jones}, C., {Murray}, S.~S., {Forman}, W.~R.,
  {Dey}, A., {Brown}, M.~J.~I., {Eisenhardt}, P., {Gonzalez}, A.~H., {Green},
  P., \& {Stern}, D. 2012, \apjs, 200, 8

\bibitem[{{Komossa} {et~al.}(2003){Komossa}, {Burwitz}, {Hasinger}, {Predehl},
  {Kaastra}, \& {Ikebe}}]{KO03.1}
{Komossa}, S., {Burwitz}, V., {Hasinger}, G., {Predehl}, P., {Kaastra}, J.~S.,
  \& {Ikebe}, Y. 2003, \apjl, 582, L15

\bibitem[{{Komossa} {et~al.}(2008){Komossa}, {Xu}, {Zhou}, {Storchi-Bergmann},
  \& {Binette}}]{KO08.1}
{Komossa}, S., {Xu}, D., {Zhou}, H., {Storchi-Bergmann}, T., \& {Binette}, L.
  2008, \apj, 680, 926

\bibitem[{{Koss} {et~al.}(2011){Koss}, {Mushotzky}, {Treister}, {Veilleux},
  {Vasudevan}, {Miller}, {Sanders}, {Schawinski}, \& {Trippe}}]{KO11.1}
{Koss}, M., {Mushotzky}, R., {Treister}, E., {Veilleux}, S., {Vasudevan}, R.,
  {Miller}, N., {Sanders}, D.~B., {Schawinski}, K., \& {Trippe}, M. 2011,
  \apjl, 735, L42+

\bibitem[{{Koss} {et~al.}(2012){Koss}, {Mushotzky}, {Treister}, {Veilleux},
  {Vasudevan}, \& {Trippe}}]{KO12.1}
{Koss}, M., {Mushotzky}, R., {Treister}, E., {Veilleux}, S., {Vasudevan}, R.,
  \& {Trippe}, M. 2012, \apjl, 746, L22

\bibitem[{{Lin} {et~al.}(2008){Lin}, {Patton}, {Koo}, {Casteels}, {Conselice},
  {Faber}, {Lotz}, {Willmer}, {Hsieh}, {Chiueh}, {Newman}, {Novak}, {Weiner},
  \& {Cooper}}]{LI08.1}
{Lin}, L., {Patton}, D.~R., {Koo}, D.~C., {Casteels}, K., {Conselice}, C.~J.,
  {Faber}, S.~M., {Lotz}, J., {Willmer}, C.~N.~A., {Hsieh}, B.~C., {Chiueh},
  T., {Newman}, J.~A., {Novak}, G.~S., {Weiner}, B.~J., \& {Cooper}, M.~C.
  2008, \apj, 681, 232

\bibitem[{{Liu} {et~al.}(2013){Liu}, {Civano}, {Shen}, {Green}, {Greene}, \&
  {Strauss}}]{LI13.1}
{Liu}, X., {Civano}, F., {Shen}, Y., {Green}, P., {Greene}, J.~E., \&
  {Strauss}, M.~A. 2013, \apj, 762, 110

\bibitem[{{Liu} {et~al.}(2010{\natexlab{a}}){Liu}, {Greene}, {Shen}, \&
  {Strauss}}]{LI10.2}
{Liu}, X., {Greene}, J.~E., {Shen}, Y., \& {Strauss}, M.~A. 2010{\natexlab{a}},
  \apjl, 715, L30

\bibitem[{{Liu} {et~al.}(2010{\natexlab{b}}){Liu}, {Shen}, {Strauss}, \&
  {Greene}}]{LI10.1}
{Liu}, X., {Shen}, Y., {Strauss}, M.~A., \& {Greene}, J.~E. 2010{\natexlab{b}},
  \apj, 708, 427

\bibitem[{{Lotz} {et~al.}(2011){Lotz}, {Jonsson}, {Cox}, {Croton}, {Primack},
  {Somerville}, \& {Stewart}}]{LO11.1}
{Lotz}, J.~M., {Jonsson}, P., {Cox}, T.~J., {Croton}, D., {Primack}, J.~R.,
  {Somerville}, R.~S., \& {Stewart}, K. 2011, \apj, 742, 103

\bibitem[{{Ludwig} {et~al.}(2012){Ludwig}, {Greene}, {Barth}, \& {Ho}}]{LU12.1}
{Ludwig}, R.~R., {Greene}, J.~E., {Barth}, A.~J., \& {Ho}, L.~C. 2012, \apj,
  756, 51

\bibitem[{{McGurk} {et~al.}(2011){McGurk}, {Max}, {Rosario}, {Shields},
  {Smith}, \& {Wright}}]{MC11.1}
{McGurk}, R.~C., {Max}, C.~E., {Rosario}, D.~J., {Shields}, G.~A., {Smith},
  K.~L., \& {Wright}, S.~A. 2011, \apjl, 738, L2

\bibitem[{{Milosavljevi{\'c}} \& {Merritt}(2001)}]{MI01.1}
{Milosavljevi{\'c}}, M., \& {Merritt}, D. 2001, \apj, 563, 34

\bibitem[{{Mullaney} \& {Ward}(2008)}]{MU08.1}
{Mullaney}, J.~R., \& {Ward}, M.~J. 2008, \mnras, 385, 53

\bibitem[{{Newman} {et~al.}(2012)}]{NE12.1}
{Newman}, J.~A., {et~al.} 2012, ArXiv e-prints

\bibitem[{{Rosario} {et~al.}(2011){Rosario}, {McGurk}, {Max}, {Shields},
  {Smith}, \& {Ammons}}]{RO11.1}
{Rosario}, D.~J., {McGurk}, R.~C., {Max}, C.~E., {Shields}, G.~A., {Smith},
  K.~L., \& {Ammons}, S.~M. 2011, \apj, 739, 44

\bibitem[{{Sarzi} {et~al.}(2006){Sarzi}, {Falc{\'o}n-Barroso}, {Davies},
  {Bacon}, {Bureau}, {Cappellari}, {de Zeeuw}, {Emsellem}, {Fathi},
  {Krajnovi{\'c}}, {Kuntschner}, {McDermid}, \& {Peletier}}]{SA06.1}
{Sarzi}, M., {Falc{\'o}n-Barroso}, J., {Davies}, R.~L., {Bacon}, R., {Bureau},
  M., {Cappellari}, M., {de Zeeuw}, P.~T., {Emsellem}, E., {Fathi}, K.,
  {Krajnovi{\'c}}, D., {Kuntschner}, H., {McDermid}, R.~M., \& {Peletier},
  R.~F. 2006, \mnras, 366, 1151

\bibitem[{{Shen} {et~al.}(2011){Shen}, {Liu}, {Greene}, \& {Strauss}}]{SH11.1}
{Shen}, Y., {Liu}, X., {Greene}, J.~E., \& {Strauss}, M.~A. 2011, \apj, 735, 48

\bibitem[{{Shen} {et~al.}(2013){Shen}, {Liu}, {Loeb}, \& {Tremaine}}]{SH13.1}
{Shen}, Y., {Liu}, X., {Loeb}, A., \& {Tremaine}, S. 2013, ArXiv e-prints

\bibitem[{{Shields}(1992)}]{SH92.1}
{Shields}, J.~C. 1992, \apjl, 399, L27

\bibitem[{{Smith} {et~al.}(2010){Smith}, {Shields}, {Bonning}, {McMullen},
  {Rosario}, \& {Salviander}}]{SM10.1}
{Smith}, K.~L., {Shields}, G.~A., {Bonning}, E.~W., {McMullen}, C.~C.,
  {Rosario}, D.~J., \& {Salviander}, S. 2010, \apj, 716, 866

\bibitem[{{Tingay} \& {Wayth}(2011)}]{TI11.1}
{Tingay}, S.~J., \& {Wayth}, R.~B. 2011, \aj, 141, 174

\bibitem[{{Van Wassenhove} {et~al.}(2012){Van Wassenhove}, {Volonteri},
  {Mayer}, {Dotti}, {Bellovary}, \& {Callegari}}]{VA12.1}
{Van Wassenhove}, S., {Volonteri}, M., {Mayer}, L., {Dotti}, M., {Bellovary},
  J., \& {Callegari}, S. 2012, \apjl, 748, L7

\bibitem[{{Veilleux} \& {Osterbrock}(1987)}]{VE87.1}
{Veilleux}, S., \& {Osterbrock}, D.~E. 1987, \apjs, 63, 295

\bibitem[{{Veilleux} {et~al.}(2001){Veilleux}, {Shopbell}, \&
  {Miller}}]{VE01.1}
{Veilleux}, S., {Shopbell}, P.~L., \& {Miller}, S.~T. 2001, \aj, 121, 198

\bibitem[{{Wang} {et~al.}(2009){Wang}, {Chen}, {Hu}, {Mao}, {Zhang}, \&
  {Bian}}]{WA09.1}
{Wang}, J., {Chen}, Y., {Hu}, C., {Mao}, W., {Zhang}, S., \& {Bian}, W. 2009,
  \apjl, 705, L76

\bibitem[{{Zamanov} {et~al.}(2002){Zamanov}, {Marziani}, {Sulentic}, {Calvani},
  {Dultzin-Hacyan}, \& {Bachev}}]{ZA02.1}
{Zamanov}, R., {Marziani}, P., {Sulentic}, J.~W., {Calvani}, M.,
  {Dultzin-Hacyan}, D., \& {Bachev}, R. 2002, \apjl, 576, L9

\end{thebibliography}

\end{document}